\documentclass[conference]{IEEEtran}
\usepackage{cite}
\usepackage{amsmath,amssymb,amsfonts}
\usepackage{algorithmic}
\usepackage{graphicx}
\usepackage{textcomp}
\usepackage{xcolor}
\usepackage{fancyhdr}
\usepackage[hyphens]{url}

\usepackage[draft]{fixme}

\usepackage{cleveref}

\def\BibTeX{{\rm B\kern-.05em{\sc i\kern-.025em b}\kern-.08em
    T\kern-.1667em\lower.7ex\hbox{E}\kern-.125emX}}

\title{Fetch-Directed Instruction Prefetching Revisited}
\author{}

\author{\IEEEauthorblockN{Truls Asheim}
\IEEEauthorblockA{NTNU, Norway\\
truls.asheim@ntnu.no}
\and
\IEEEauthorblockN{Rakesh Kumar}
\IEEEauthorblockA{NTNU, Norway\\
rakesh.kumar@ntnu.no}
\and
\IEEEauthorblockN{Boris Grot}
\IEEEauthorblockA{University of Edinburgh, UK\\
boris.grot@ed.ac.uk}
}

\begin{document}
\maketitle
\pagestyle{plain}


\begin{abstract}
  Prior work has observed that fetch-directed prefetching (FDIP) is highly effective at covering instruction cache misses. The key to FDIP's effectiveness is having a sufficiently large BTB to accommodate the application's branch working set. In this work, we introduce several optimizations that significantly extend the reach of the BTB within the available storage budget. Our optimizations target nearly every source of storage overhead in each BTB entry; namely, the tag, target address, and size fields. 
  
  We observe that while most dynamic branch instances have short offsets, a large number of branches has longer offsets or requires the use of full target addresses. Based on this insight, we break-up the BTB into multiple smaller BTBs, each storing offsets of different length. This enables a dramatic reduction in storage for target addresses. We further compress tags to 16 bits and avoid the use of the basic-block-oriented BTB advocated in prior FDIP variants. The latter optimization eliminates the need to store the basic block size in each BTB entry. Our final design, called FDIP-X, uses an ensemble of 4 BTBs and always outperforms conventional FDIP with a unified basic-block-oriented BTB for equal storage budgets. 
\end{abstract} 

\section{Introduction}

Contemporary server applications feature deeply layered software
stacks and massive instruction working sets far exceeding the capacity
of the instruction cache found in current processors. This causes a
large number of front-end stall cycles due to frequent instruction cache misses, motivating research into front-end prefetching. 

Existing research on instruction prefetching has introduced several schemes that use the BTB and the branch predictor to drive front-end prefetching. These schemes effectively allow the instruction address generation logic to run ahead of the actual fetch stream by predicting and resolving future branches, and issuing prefetches to the generated candidate addresses. This idea, called Fetch Directed Instruction Prefetching (FDIP), was pioneered by Reinman et al.~\cite{reinman99}. Subsequent works extended FDIP to also prefetch into the branch target buffer (BTB)~\cite{boomerang}, and, most recently, introduced a compressed BTB design to maximize instruction footprint coverage with a limited BTB storage budget~\cite{shotgun}. 

A key conclusion reached by recent front-end prefetching research is that FDIP is highly effective at covering L1-I misses, and -- given a sufficiently large BTB -- is competitive with storage-intensive front-end prefetchers such as temporal streaming~\cite{tifs, pif, shift}.
The reason why the BTB plays a key role in FDIP and its derivates is that the BTB is used to identify branches, which -- if taken -- redirect the control flow to a {\em target address}. For each branch tracked in the BTB, its target address is also stored there.  Thus, the number of branches tracked in BTB plays a key role in FDIP's effectiveness because any branch evicted from the BTB (e.g., due to contention) may lower FDIP's effectiveness by impeding its ability to identify control flow discontinuities. 

However, naively increasing the BTB size to track more branches results in massive storage overhead. Therefore, this work focuses on optimizing the BTB entry organization to reduce its storage requirements; thus maximizing number of branches captured in a given storage budget.

Our proposed BTB organization leverages the insight that branch offset lengths are unequally distributed. Conditional branches have shorter offsets than unconditional branches, which require full target addresses. Moreover, within the conditional branch category, many targets are very close to the branch itself, requiring very few bits to encode the offset. 
Based on this insight, we partition the BTB into several smaller BTBs, each storing branches whose targets fall within a certain distance of the branch itself. Because the target field accounts for over half of each entry's storage budget in a baseline BTB design (see Figure~\ref{fig:conv-btb}), this optimization brings significant storage savings. 

We further observe that a full tag is not necessary to identify branches and, instead, use a shorter, hashed tag. Through empirical studies, we find that a 16-bit tag achieves a significant reduction in storage cost compared to the full 39-bit tag\footnote{Assuming a 48-bit virtual address space, 128-set BTB, and word (32-bit) aligned instructions.} with negligible performance impact. 

Lastly, we eschew a basic-block-oriented BTB used in all prior FDIP-based designs and opt for a conventional BTB. This provides further storage savings by avoiding the need to track basic block size in each BTB entry. While this optimization carries no performance cost, in practice it may increase BTB bandwidth requirements and its power consumption. 

Our final FDIP design, called {\em FDIP-X}, uses 4 separate BTBs with each containing only 16-bit hashed tags and no basic block size information. The 4 BTBs only differ in the number of bits they allocate to store branch target offsets. Our evaluation shows that FDIP-X significantly outperforms FDIP under stringent storage budget on both server and client traces. However, when the storage budget restriction are relaxed both designs perform similar. 
\section{FDIP Basics}
\label{sec:background}

\begin{figure}
\includegraphics[width=\linewidth]{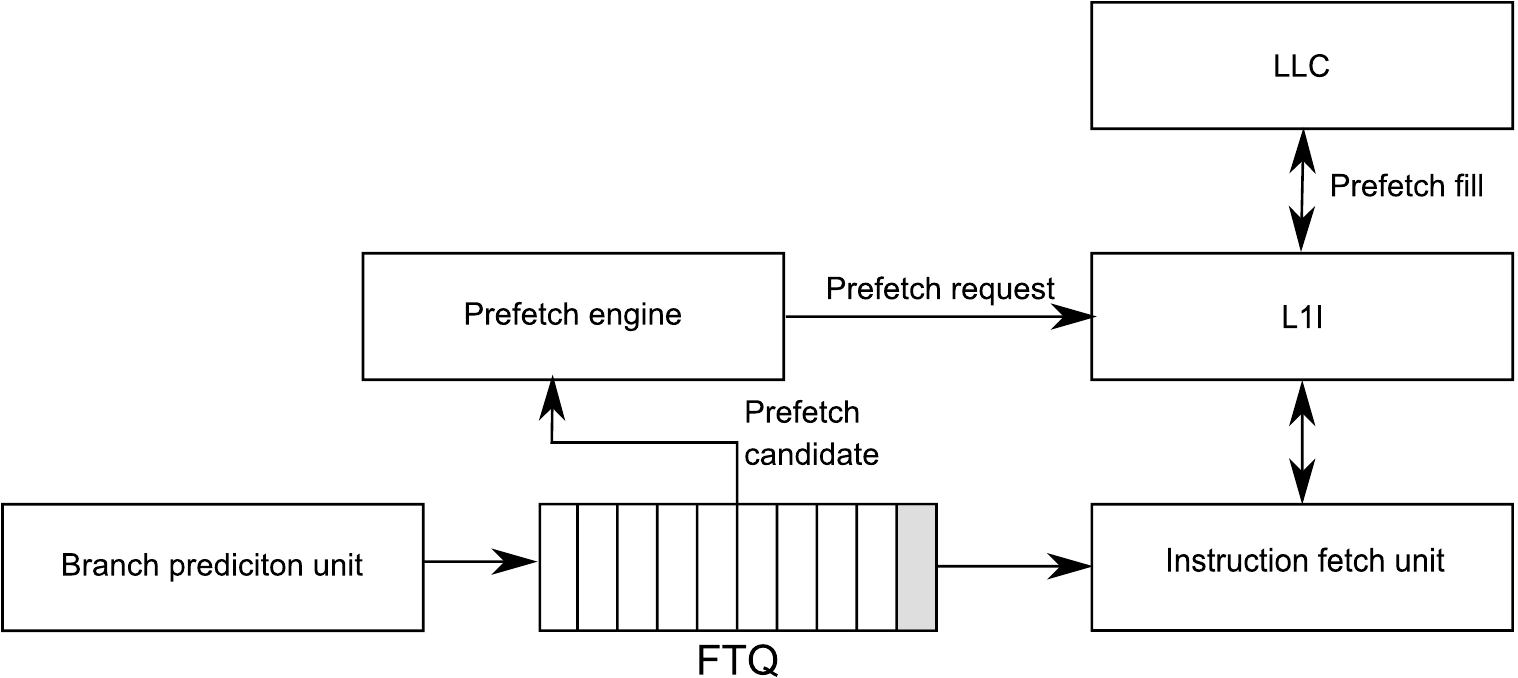}
\caption{The FDIP microarchitecture}
\label{fig:fdip}
\end{figure}

The baseline for this implementation is fetch directed instruction
prefetch, FDIP \cite{reinman99}. FDIP, sketched in \Cref{fig:fdip}, is an instruction prefetching
method that predicts future control flow based on the information
contained within the branch-prediction unit, encompassing the branch
predictor, BTB, and return address stack. The key innovation pioneered by FDIP is
the decoupling of the branch-prediction unit and the fetch engine via the {\em fetch target queue (FTQ)}. This decoupling allows the branch
prediction unit to run ahead of the fetch engine and predict future control flow. The head of the FTQ, shaded in \Cref{fig:fdip}, is the fetch point, while subsequent entries can be used for issuing prefetches as described below. 

The original FDIP proposal relies on a basic block-oriented BTB, 
which stores the start and length of basic blocks rather than branch
instruction addresses. Here, a basic block is defined as straight-line
code ending in a branch instruction. On every cycle, the branch prediction unit predicts the next basic
block and inserts it into the FTQ. In this way, the FTQ contains a
stream of predicted basic blocks to be fetched. An FTQ entry contains
information about the basic block corresponding to the current
fetch entry. The head of the FTQ is consumed by the fetch engine which
issues N demand-fetch requests where N is the fetch-width.

Since the non-head entries of the FTQ contain addresses that will be fetched
by the fetch engine in the future, they represent ideal prefetch
candidates. The prefetch engine is the component responsible for
scanning the contents of the FTQ to look for new prefetch
candidates. For every candidate discovered, the prefetch engine issues
a prefetch probe, which checks if the L1-I block corresponding to the
FTQ entry is present in the L1-I. If not, FDIP issues a prefetch
request to bring the block from higher cache levels into
the L1-I. Requests to the L1-I are prioritized
such that demand fetches from the fetch engine are processed before
prefetch probes.

\section{FDIP-X}
\label{sec:design}

FDIP-X deploys several optimizations, all aimed at maximizing BTB reach. 
To motivate the design, we refer the reader to Figure~\ref{fig:conv-btb}, which shows a conventional BTB and the composition of each entry. The following sections describe optimizations aimed at reducing or eliminating the storage cost of the three costliest fields making up each BTB entry: offset/target, tag, BB size.

\subsection{Partitioned BTB}

\begin{figure}
\centering
\includegraphics[width=0.9\linewidth]{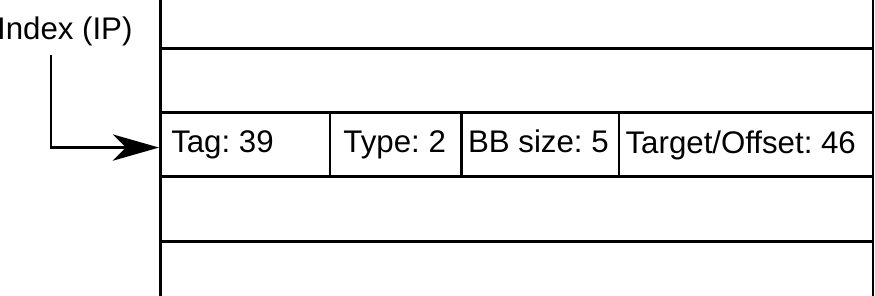}
\caption {The composition of entries in a basic-block-based BTB. The numbers are the number of bits used to encode each field.}
\label{fig:conv-btb}
\end{figure}

\begin{figure*}[t!]
    \centering
    \includegraphics[width=0.8\textwidth, trim=60 180 60 155, clip]{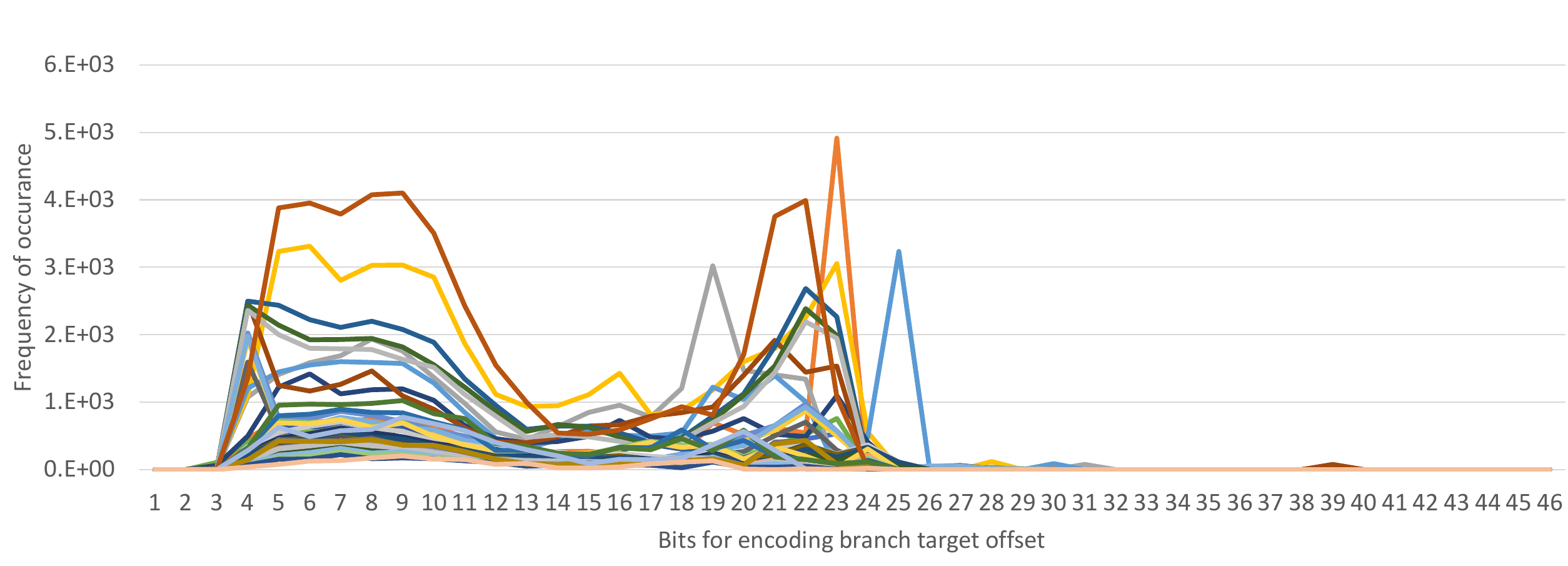}
    \caption{Distribution of branch target offsets.}
    \label{fig:offsets}
\end{figure*}

As Figure~\ref{fig:conv-btb} shows, the single largest contributor to storage cost is the offset/target field, which stores the branch offset or the target address -- up to 46 bits long. Our key insight is that most branches use offsets shorter than 46 bits. Figure~\ref{fig:offsets} plots the distribution of offsets\footnote{In our traces, all the instructions are word (32-bit) aligned as the traces are generated on ARMv8. Therefore, the branch target offset is calculated as the distance to target in instructions rather than in bytes.} in the branch working sets of our workload traces. The X-axis shows the number of bits required to encode the offset, while the Y-axis plots the frequency with which the given offset size occurs in each trace. Note that, in addition to bits for encoding the offset, an additional bit is required for the direction of the offset (forward/backward). 

As the figure shows, shorts offsets dominate. Indeed, very few branches have an offset requiring more than 23 bits to encode. Note that the data includes both conditional branches and unconditional jumps, hence it comprehensively covers the full branch working set for these traces. 

Based on the insights gleaned from Figure~\ref{fig:offsets}, we propose to partition a single logical BTB into multiple physically-separate BTBs. The BTBs differ amongst themselves only in the size of the offset/target field. When the branch prediction unit queries an address, all BTB partitions are queried in parallel, hence presenting a logical equivalent of a monolithic BTB. 

\begin{figure}
    \centering
    \includegraphics[width=0.8\linewidth]{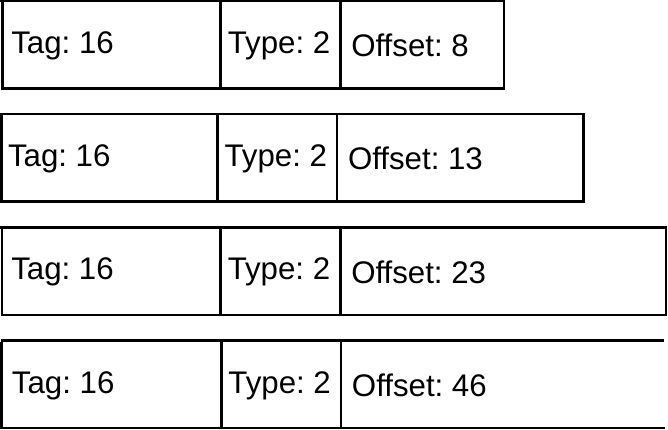}
    \caption{The FDIP-X BTB organization.} 
    \label{fig:btb-dist}
\end{figure}

\Cref{fig:btb-dist} shows the partitioning used in this implementation. We use four different BTBs with offset field sizes of 8-bits, 13-bits, 23-bits and 46-bits. Branches are allocated entries in one of these BTBs based on the minimum number of bits required to encode their target offsets. For example, if a branch requires 10 bits for encoding its target offset, it is allocated an entry in the BTB with target offset field of 13-bits. 

We also leverage the insights from Figure~\ref{fig:offsets} to size the different BTBs. For example, as very few branches require more than 23 bits to encode their target offsets, the BTB with 46-bit offset field is allocated the least number of entries. Also, the remaining three BTBs (8-, 13-, and 23-bit offset) are allocated similar number of entries, as the frequency of 0-8 bit, 9-13 bit, and 14-23 bit offsets is about same.

\subsection{Tag compression}

Tags comprise the second largest source of storage overhead in each BTB entry, requiring 39 bits in the baseline design. 
To further reduce the BTB storage requirement, FDIP-X uses a compressed 16-bit tag in all of its BTBs. Our compression scheme maintains the 8 low-order bits same as in the full tag. The remaining bits of the full tag are folded, using the XOR operator, in blocks of 8 to get the 8 higher-order bits for the compressed tag. The performance impact of this scheme is negligible as the hashing function (folded XOR) preserves most of the entropy found in the high-order bits.

\subsection{Block based or conventional BTB?}
Prior incarnations of FDIP used some variant of a block-based BTB as described in Section~\ref{sec:background}.
The advantage of using such a BTB is that each entry contains the location of the next branch. This reduces the number of times the BTB has to be queried to locate the next branch, which saves BTB bandwidth and power. The disadvantage of a block-based BTB, however, is that each BTB entry needs to store the size of the associated basic block. 

To reduce BTB storage requirements, FDIP-X deploys a conventional instruction-based BTB. Such a BTB is accessed with an instruction address and a hit in the BTB indicates that the address corresponds to a branch instruction. In addition, BTB provides information about the branch type (conditional, call, etc.) and the target of the branch. If a branch is predicted to be taken (or in the case of an unconditional branches), address generation resumes from the branch target. If a branch is not found in a BTB, addresses continue to be generated sequentially. 

The space saving of an instruction-based BTB (compared to a block-based one) is directly proportional to the number of entries in the BTB. For example, with a 8K-entry BTB, not needing the block size field (which requires 5 bits per entry) saves 5KB of storage. Furthermore, empirically, we could not observe a performance difference between the two BTB organizations.

\subsection{Prefetch Throttling}

A key design parameter of an instruction prefetcher is prefetch throttling. Uncontrolled wrong-path prefetching can be detrimental for performance as it wastes on-chip bandwidth and may evict useful instructions from the instruction cache. 

FDIP-X throttles prefetches by maintaining a list of recently issued prefetches. This list is used to filter the prefetch requests by suppressing the prefetches for cache blocks that have been recently requested. FDIP-X uses a 10-entry fully-associative table to track recently issued prefetches. 

In addition, the FTQ itself acts as a throttling mechanism. This is because no new addresses can be generated once the FTQ is full.

\section{Evaluation}
\label{sec:eval}

We evaluate FDIP-X on the traces and simulation infrastructure provided by IPC-1~\cite{ipc1}. First, we breakdown the BTB storage requirements, followed by a performance comparison between FDIP-X and FDIP for different BTB storage budgets. Our performance comparison also includes PIF~\cite{pif}, a state-of-the-art temporal stream prefetcher. Finally, we evaluate the performance impact of tag compression.

\subsection{Storage break-down}

\begin{scriptsize}
\begin{table}[h!]
  \centering
  \caption{Storage breakdown for basic-block-oriented BTB}
  \label{table:metadata}
  \begin{tabular}{lrrr} \hline
    \textbf{Entries} & \textbf{Organization} & \textbf{Entry size (bits)}
    & \textbf{Total (bytes)} \\\hline
    1K & 128-set, 8-way & 92 & 11.5K\\\hline
    2K & 256-set, 8-way & 91 & 22.75K\\\hline
    4K & 512-set, 8-way & 90 & 45K\\\hline
    8K & 1024-set, 8-way & 89 & 89K\\\hline
    16K & 2048-set, 8-way & 88 & 176K\\\hline
    32K & 4096-set, 8-way & 87 & 348K \\\hline
  \end{tabular}
\end{table}
\end{scriptsize}

\begin{scriptsize}
\begin{table}[h!]
  \centering
  \caption{Storage breakdown for FDIP-X BTB}
  \label{table:metadata-fdipx}
  \begin{tabular}{lcr} \hline
    \textbf{Budget}&\textbf{Distribution}&\textbf{Used}\\
    \textbf{(KB)}&&\textbf{(KB)}\\\hline
    11.5&
    \begin{tabular}{lrrr} 
    \textbf{BTB}&\textbf{Entry size}&\textbf{Entries}&\textbf{Storage}\\\hline
    8-bit offset&26-bit&768&2.44KB\\
    13-bit offset&31-bit&768&2.9KB\\
    23-bit offset&41-bit&768&3.84KB \\
    46-bit offset&64-bit&112& 0.88KB\\
    \end{tabular}&10.06\\\hline
    
    22.75&
    \begin{tabular}{lrrr} 
    \textbf{BTB}&\textbf{Entry size}&\textbf{Entries}&\textbf{Storage}\\\hline
    8-bit offset&26-bit&1.5K&4.88KB\\
    13-bit offset&31-bit&1.5K&5.81KB\\
    23-bit offset&41-bit&1.5K&7.68KB\\
    46-bit offset&64-bit&224&1.75KB\\
    \end{tabular}&20.12\\\hline
    
    45&
    \begin{tabular}{lrrr} 
    \textbf{BTB}&\textbf{Entry size}&\textbf{Entries}&\textbf{Storage}\\\hline
    8-bit offset&26-bit&3K&9.75KB\\
    13-bit offset&31-bit&3K&11.63KB\\
    23-bit offset&41-bit&3K&15.37KB\\
    46-bit offset&64-bit&448&3.5KB\\
    \end{tabular}&40.25\\\hline
    
    89&
    \begin{tabular}{lrrr} 
    \textbf{BTB}&\textbf{Entry size}&\textbf{Entries}&\textbf{Storage}\\\hline
    8-bit offset&26-bit&6K&19.5KB\\
    13-bit offset&31-bit&6K&23.25KB\\
    23-bit offset&41-bit&6K&30.75KB\\
    46-bit offset&64-bit&896&7KB\\
    \end{tabular}&80.5\\\hline
    
    176&
    \begin{tabular}{lrrr} 
    \textbf{BTB}&\textbf{Entry size}&\textbf{Entries}&\textbf{Storage}\\\hline
    8-bit offset&26-bit&12K&39KB\\
    13-bit offset&31-bit&12K&46.5KB\\
    23-bit offset&41-bit&12K&61.5KB\\
    46-bit offset&64-bit&1.75K&14KB\\
    \end{tabular}&161\\\hline
    
    348&
    \begin{tabular}{lrrr} 
    \textbf{BTB}&\textbf{Entry size}&\textbf{Entries}&\textbf{Storage}\\\hline
    8-bit offset&26-bit&24K&78KB\\
    13-bit offset&31-bit&24K&93KB\\
    23-bit offset&41-bit&24K&123KB\\
    46-bit offset&64-bit&3.5K&28KB\\
    \end{tabular}&322\\\hline
    
  \end{tabular}
\end{table}
\end{scriptsize}

The storage requirements for a conventional basic-block-oriented BTB for different number of BTB entries are presented in  
\Cref{table:metadata} assuming a 48-bit virtual address space. We increase the number of sets in the BTB to increase the number of entries while keeping the associativity same (8-way). Notice that the entry size reduces by one bit while doubling the number of entries. This is because the size of tag reduces as more bits are needed to index the BTB.

\Cref{table:metadata-fdipx} presents the distribution of storage budget of a basic-block-oriented BTB among different BTBs (8-bit, 13-bit, 23-bit, and 46-bit offsets) in FDIP-X. Like basic-block-oriented BTB, we double the number of sets to double the BTB capacity while maintaining the associativity (6-way). Also notice that since the number of sets have to be a power of 2, we are not able to precisely match the storage of basic-block-oriented BTB and FDIP-X BTB. In fact, basic-block-oriented BTB gets a higher storage budget especially with more entries. Yet, FDIP-X BTBs together provide about 2.36x entries than basic-block-oriented BTB.

\subsection{Performance}

\begin{figure}[t!]
    \centering
    \includegraphics[width=\columnwidth, trim=35 260 70 300, clip]{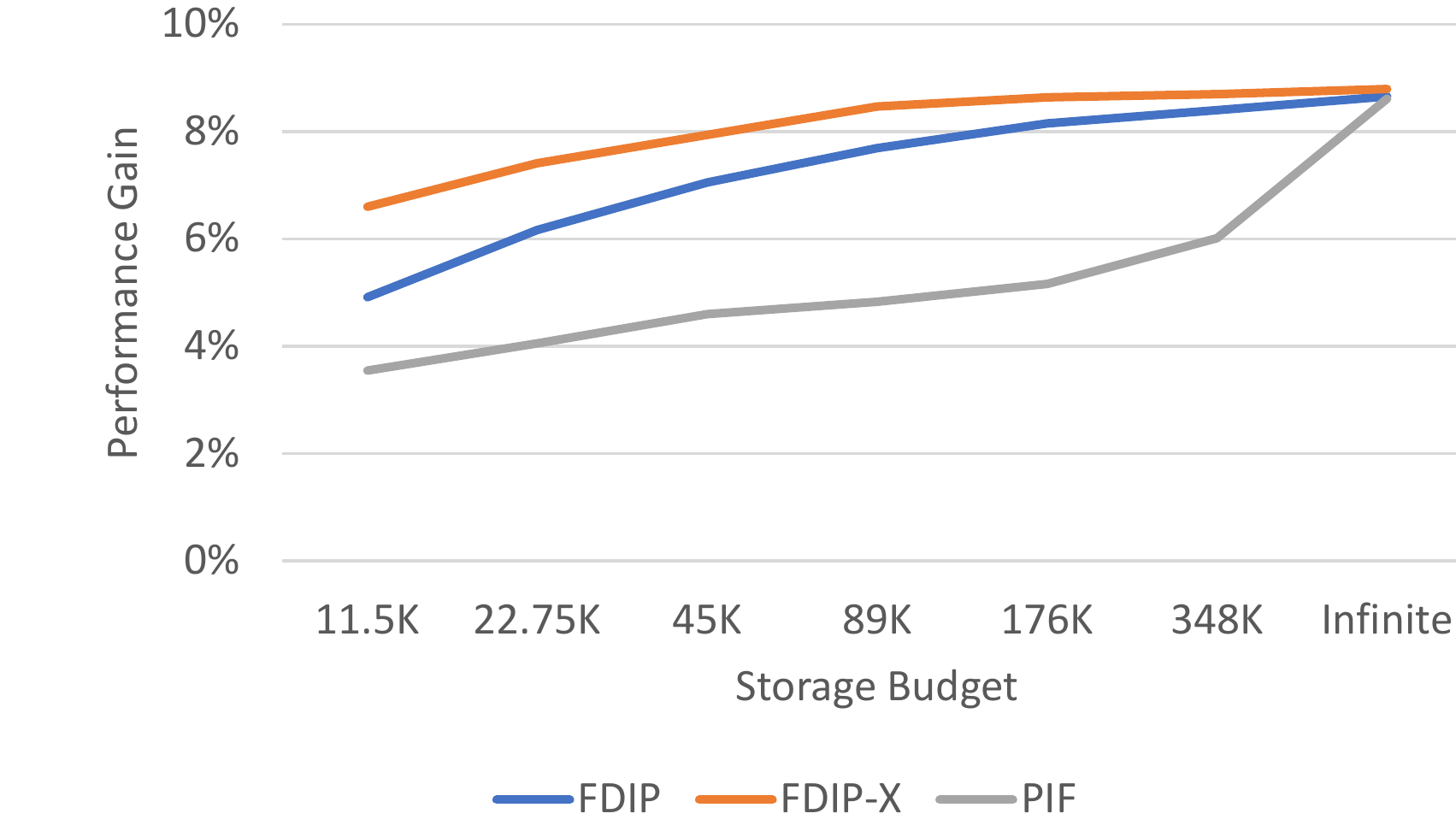}
    \caption{FDIP, FDIP-X, and PIF performance gain, over no-prefetch baseline, across \textbf{client} traces. X-axis is storage budget for a 1K-, 2K-, 4K-, 8K-, 16K-, 32K-, and infinite-entry basic-block-oriented BTB.}
    \label{fig:clientPerf}
\end{figure}

\begin{figure}[t!]
    \centering
    \includegraphics[width=\columnwidth, trim=35 260 70 300, clip]{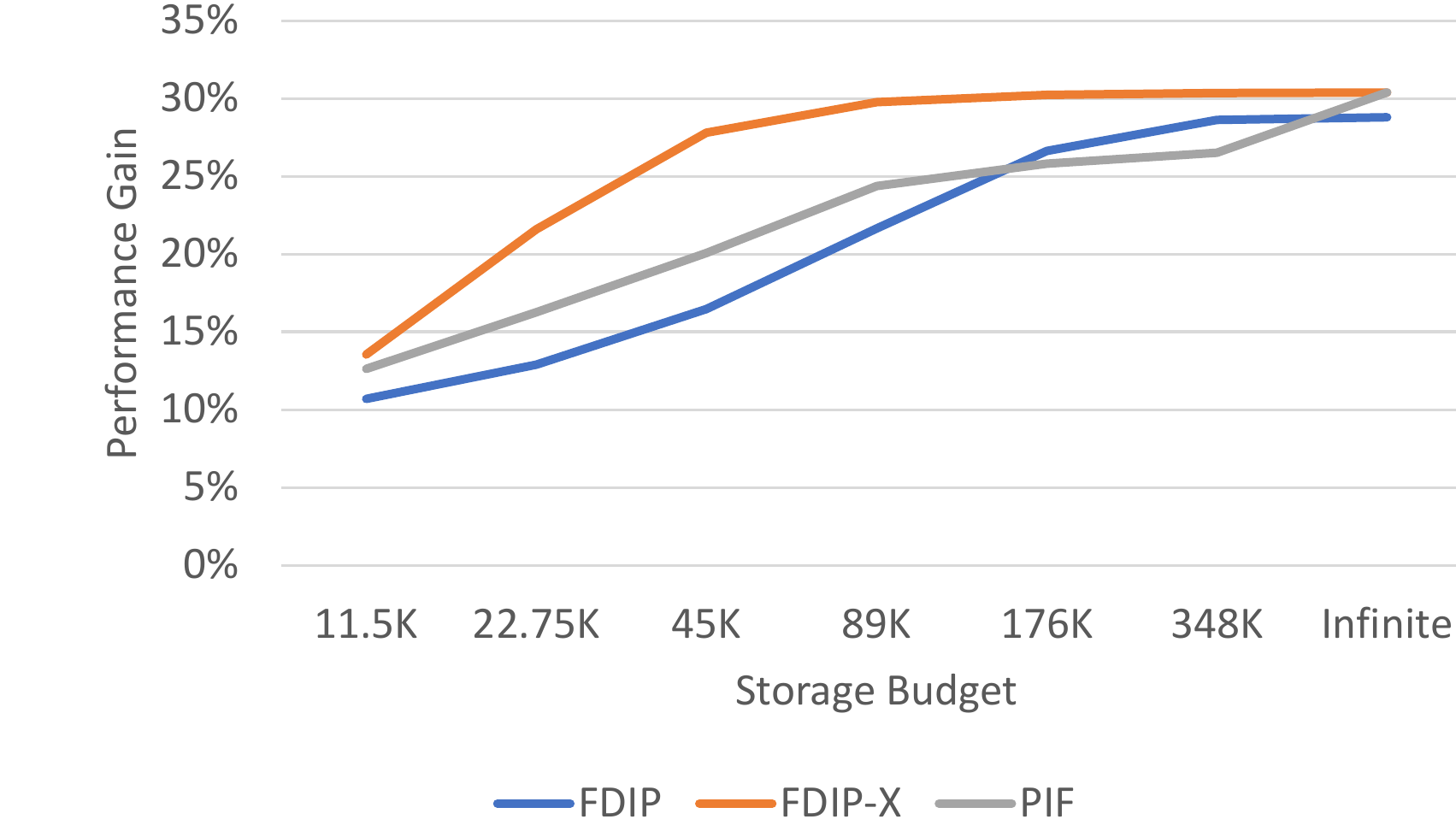}
    \caption{FDIP, FDIP-X, and PIF performance gain, over no-prefetch baseline, across \textbf{server} traces. X-axis is storage budget for a 1K-, 2K-, 4K-, 8K-, 16K-, 32K-, and infinite-entry basic-block-oriented BTB.}
    \label{fig:serverPerf}
\end{figure}

Figures~\ref{fig:clientPerf} and~\ref{fig:serverPerf} compare the performance gains of FDIP, FDIP-X, and PIF, over a no-prefetch baseline, for different storage budgets across client and server traces respectively. The storage budgets correspond to 1K-, 2K-, 4K-, 8K-, 16K-, 32K-, and infinite-entry basic-block-oriented BTB. 

\begin{figure*}[t!]
    \centering
    \includegraphics[width=\textwidth, trim=100 190 55 200, clip]{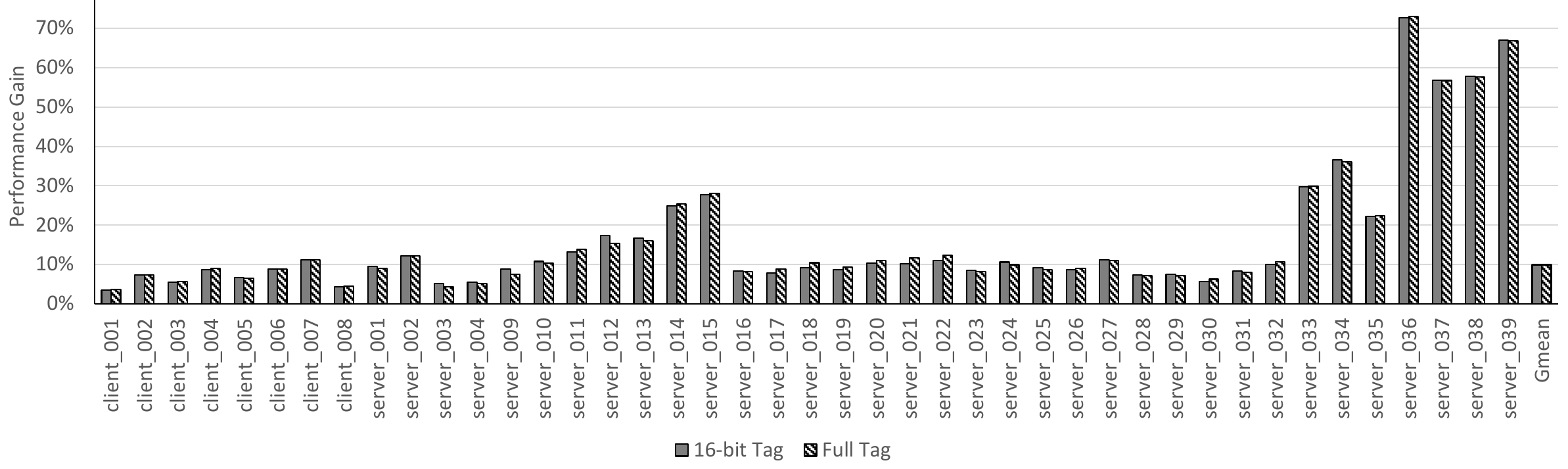}
    \caption{FDIP-X performance gain with 16-bit tags and full tags in BTB.}
    \label{fig:FullTagperf}
\end{figure*}

As the figures show, FDIP-X comprehensively outperforms FDIP and PIF for practical storage budgets of few tens of kilobytes. The performance advantage of FDIP-X is especially visible on server traces as they put high pressure on instruction cache due to their massive instruction footprints. As Figure~\ref{fig:serverPerf} shows FDIP-X needs only 45KB of storage to reach within 2.5\% of the performance offered by an infinite BTB (27.8\% vs 30.3\%). In contrast, FDIP with its basic-block-oriented BTB requires nearly 200KB of storage to reach similar performance level. Similarly, PIF also requires significantly higher storage budget than FDIP-X to deliver similar performance in a  practical storage budget range.

The figures also show that the client traces offer much less performance opportunity compared to the server traces due to their smaller instruction footprints. As a result, the performance gap between FDIP-X and FDIP narrows down quickly as the BTB storage budget increases. PIF, in contrast, falls significantly behind both FDIP-X and FDIP at practical storage budgets.

Figures~\ref{fig:clientPerf} and~\ref{fig:serverPerf} also show that, for storage budgets of up to 89KB, PIF outperforms FDIP on server traces; however, it lags behind FDIP on client traces. This is because the client workloads feature shorter streams (that are used by PIF for prefetching) which causes PIF to reset often and loose performance. Our analysis shows that PIF experiences 1.5x more resets on client traces than on the server ones.

\subsection{Impact of tag compression}

For assessing the performance loss due to compressed tags, we compare FDIP-X performance with 16-bit tags to full tags for the smallest BTB size. We choose the smallest BTB as it would suffer highest aliasing because of tag compression. As the results presented in Figure~\ref{fig:FullTagperf} show, full tags provide 9.96\% performance gain over the baseline compared to 9.92\% with compressed tags, a difference of only 0.04\%. This result shows that our tag compression mechanism is able to preserve the entropy of the higher order bits.

\vspace{.1in}

Overall, the results presented in this section show that by partitioning the BTB into several smaller BTBs, compressing tags, and avoiding the use of a block-based BTB, FDIP-X drastically increases in number of entries in a given BTB storage budget. This enables FDIP-X to deliver much higher performance than the conventional FDIP especially with stringent storage budgets.



\bibliographystyle{IEEEtranS}
\bibliography{refs}

\end{document}